\documentclass[reprint,amsmath,amssymb,apr,aip,onecolumn]{revtex4-2}
\usepackage{graphicx}
\usepackage{graphics}
\usepackage{algorithm,algorithmic}
\usepackage{mathptmx}
\usepackage{times}
\usepackage{amsmath}
\usepackage{amssymb}
\usepackage{dcolumn}
\usepackage{color}
\usepackage{bm}
\usepackage{units}
\usepackage{booktabs}
\usepackage{multirow}
\usepackage[version=4]{mhchem}
\usepackage{siunitx}[=v2]
\usepackage{charter}
\usepackage[acronym]{glossaries}

\newacronym{imc}{IMC}{in-memory computing}
\newacronym{pcm}{PCM}{phase-change memory}

\begin{document}
\title{Electro-Thermal Origins of Enhanced Performance and Scalability in Disc-Type Phase-Change Memory}

\author{Ghazi Sarwat Syed}\affiliation{IBM Research -- Europe, S\"{a}umerstrasse 4, 8803 R\"{u}schlikon, Switzerland} 
\author{Loris Coccia} \affiliation{IBM Research -- Europe, S\"{a}umerstrasse 4, 8803 R\"{u}schlikon, Switzerland}
\author{Siddharth Gautam}
\affiliation{IBM Research -- Europe, S\"{a}umerstrasse 4, 8803 R\"{u}schlikon, Switzerland}
\affiliation{École polytechnique fédérale de Lausanne (EPFL), Lausanne, Switzerland}
\author{Vara Prasad Jonnalagadda}\affiliation{IBM Research -- Europe, S\"{a}umerstrasse 4, 8803 R\"{u}schlikon, Switzerland}
\author {Abu Sebastian} \affiliation{IBM Research -- Europe, S\"{a}umerstrasse 4, 8803 R\"{u}schlikon, Switzerland} %

\maketitle

\noindent \textbf{Disc-type phase-change memory is a promising device architecture for analog in-memory computing, enabled by the unique physical properties of ultra-thin phase-change materials and the device geometry. In this study, we develop a comparative analytical framework using compact electro-thermal models to examine the state-dependent readout and programming behavior of these devices. The framework provides insight into the fundamental physical mechanisms responsible for their enhanced functionality and highlights promising directions for further device optimization and performance improvement.}

\begin{flushleft}
\end{flushleft}

\section*{Introduction}

\noindent One of the key challenges faced by AI algorithms is that of the energy consumption and the latency associated with accessing large volumes of synaptic weight data from memory. Analog \gls{imc} addresses this challenge by performing compute operations directly on the network weights stored in the non volatile states of memory devices organized in crossbar arrays on chip. The most advanced analog in-memory computing chips demonstrating the feasibility of this approach are arguably based on \gls{pcm} devices~\cite{khaddam2021hermes,Y2023legalloNatElec, Y2023ambrogioNature, syed2023memory, tsmcpcm, baldo2025aimc_gegst}.  The performance of analog \gls{imc} is strongly linked to the ability to scale-up array sizes, in order to enable a high degree of parallelism in linear computations. However, this scaling introduces critical device level constraints\cite{syed2025chemicalreviews}. First, it is necessary to increase the resistance of the device across all programmable phase configurations while maintaining a sufficiently large resistance window. Higher resistance levels reduce energy consumption during matrix vector multiplication operations and reduce non idealities such as IR drop arising from the parasitic resistance of interconnects. Second, reducing the write (programming) currents enable the use of smaller selector devices, which in turn allows for reduced unit cell area and higher integration density.\\

\begin{figure}
  \includegraphics[width=0.9\linewidth]{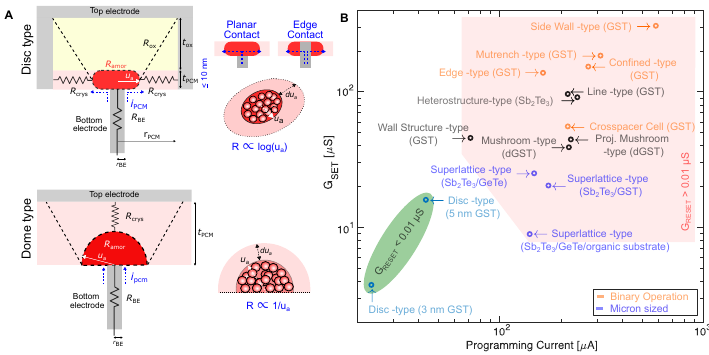}
  \caption{\textbf{Disc-type Phase Change Memory}. (A) The top panel shows a schematic cross-sectional view of a disc-type device. The device features a vertical structure, with an ultra-thin ($\leq$ \unit[10]{nm}) phase-change material layer positioned between the dielectric and bottom electrode. Top contact is made laterally at the outer edge of the phase-change layer. Based on the fabrication design, the bottom electrode can form either a planar contact or an edge contact with the phase-change material. In both cases, the current flows radially through the device. Compared to 3D dome-type configurations in prototypical devices, such as the mushroom cell (bottom panel), the amorphous regions form spatially confined cylindrical discs during RESET programming. Key geometric parameters and electrical and thermal resistances are highlighted in both panels. (B) The disc-type devices are compared with different PCM devices reported in the literature in terms of the programming current (corresponding to the onset of melting) and the SET and RESET conductance values, which are the critical device parameters for \gls{imc} application.}
\label{figure:fig1}
\end{figure}

To that end, a key device-level advance was recently introduced in the form of disc-type phase-change memory\cite{syedDisc2025} (\textit{disc-type PCM}). Disc-type PCM leverages a design strategy that minimizes both the contact area and the active volume (see Figure~\ref{figure:fig1}A) in the device. In Figure~\ref{figure:fig1}B, we compare disc-type devices with established PCM architectures\cite{Y2003haVLSI,Y2022khanNanoletters,Y2011simpsonNatureNano,Y2021sarwatAFM,Y2006pellizzerVLSI,Y2019dingScience,Y2008imIEDM,Y2007chenIEDM}. The comparison focuses on key parameters such as SET and RESET conductance and programming current, which directly impact the array sizes (also see \textcolor{black}{Supplementary Figure S1}). Notably, disc-type devices enable low programming currents and power consumption with low absolute conductance values, while remaining compatible with conventional fabrication processes and device dimensions. This makes them well suited for \gls{imc} applications. In this work, we use electro-thermal modeling to develop a comprehensive understanding of the state-dependent characteristics of disc-type devices, for which a fundamental understanding is currently lacking. Utilizing the experimental dataset presented in Ref.~\cite{syedDisc2025}, we systematically analyze the core physical mechanisms governing the device read-out characteristics and programming dynamics, specifically tracking amorphization (RESET), crystallization (SET), and endurance behaviors. Throughout, we provide an analytical comparison with prototypical PCM devices, such as mushroom\cite{Y2021sarwatAFM} and wall-structure\cite{BoniardiIEDM2014, russo2008modeling} geometries, that are characterized by a \textit{dome-type} phase configuration of the amorphous volume. 

\section*{Understanding the Read-Out Characteristics}    

\noindent The resistance state of the disc-type device, of total radius $r_\mathrm{PCM}$, is governed by the phase configuration, namely the amount and geometry of the amorphous region embedded within an otherwise crystalline phase-change material. Upon application of a short, high-current RESET pulse, a molten region forms around the bottom electrode (BE) and subsequently quenches into the amorphous phase, consistent with the electro-thermal FEM simulations shown in \textcolor{black}{Supplementary Figure S2}. The extent of this region is described by the amorphous radius $u_{\mathrm{a}}$, which ranges from the bottom-electrode radius $r_{\mathrm{BE}}$ to the phase-change-material radius $r_{\mathrm{PCM}}$. This description applies to the configuration in which the bottom electrode penetrates the phase-change material and forms an edge contact with it (Figure~\ref{figure:fig1}A).  We now discuss the characteristics of such configuration. \\

The total device resistance can be interpreted as a series combination of two contributions: the amorphous resistance $R_{\mathrm{amor}}$ from $r_{\mathrm{BE}}$ to $u_{\mathrm{a}}$, and the crystalline resistance $R_{\mathrm{crys}}$ from $u_{\mathrm{a}}$ to $r_{\mathrm{PCM}}$. 
By neglecting the small device thickness ($t_{\mathrm{PCM}}\leq 10$ nm), the system reduces to a two-dimensional geometry.  Therefore, the material is radially inhomogeneous, with conductivity $\sigma(r)=\sigma_{\mathrm{amor}}$ for $r_{\mathrm{BE}} < r \leq u_{\mathrm{a}}$ and $\sigma(r)=\sigma_{\mathrm{crys}}$ for $u_{\mathrm{a}} < r \leq r_{\mathrm{PCM}}$. Under steady-state read-out conditions and neglecting charge accumulation in the phase-change material, the electric potential satisfies the current-continuity equation $\nabla \cdot [\sigma(r)\nabla V]=0$.  Using Ohm’s law and integrating the radial current density over a cylindrical surface of height $t_{\mathrm{PCM}}$, the total current is obtained. This leads to the resistances of the two regions to follow the geometrical form,

\begin{eqnarray*}\label{eq:read-out_Rdisc}
R_{\mathrm{amor}}(u_\text{a},T,t) &=& \left[\frac{\rho_{\mathrm{amor},0}}{2 \pi t_{\mathrm{PCM}}} \ln\left(\frac{u_\text{a}}{r_{\mathrm{BE}}}\right)\right] \exp\left(\frac{E_{\mathrm{amor}}}{k_{\mathrm{B}}T}\right) \left(\frac{t}{t_0}\right)^{\nu_{\mathrm{amor}}} \\
R_{\mathrm{crys}}(u_\text{a},T,t) &=& \left[\frac{\rho_{\mathrm{crys},0}}{2 \pi t_{\mathrm{PCM}}} \ln\left(\frac{r_{\mathrm{PCM}}}{u_\text{a}}\right)\right] \exp\left(\frac{E_{\mathrm{crys}}}{k_{\mathrm{B}}T}\right) \left(\frac{t}{t_0}\right)^{\nu_{\mathrm{crys}}}
\end{eqnarray*}

Here, $\rho_{\mathrm{amor},0}$ and $\rho_{\mathrm{crys},0}$ are the temperature-independent resistivity prefactors ($1/\sigma_\text{crys/amor,0}$) at the reference time $t_0$. The expressions can further incorporate both temperature dependence through the activation energies $E_{\mathrm{amor}}$ and $E_{\mathrm{crys}}$, and temporal drift through the exponents $\nu_{\mathrm{amor}}$ and $\nu_{\mathrm{crys}}$. 

\begin{figure}[h!]
    \includegraphics[width=0.65\textwidth]{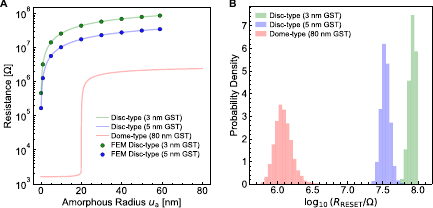}
    \caption{\textbf{Read Characteristics}. (A) The device resistance as a function of the amorphous extent ($u_\text{a}$) for disc-type and dome-type geometries, computed using the analytical model. In disc-type devices, current always traverses the amorphous region regardless of its size, whereas in dome-type geometries this occurs only once the amorphous volume fully covers the bottom electrode (for $u_\text{a} \geq r_\text{b.e}$). The scatter points represent resistance values obtained using finite-element simulations, showing a good agreement with the analytical model. (B) Variation in RESET resistance due to fabrication-induced variability in the bottom-electrode radius, $r_{\mathrm{BE}}$. The radius is sampled from a Gaussian distribution with mean $20~\mathrm{nm}$ and standard deviation $20\,\%$ of the mean, while the amorphous mark size is fixed at $u_{\mathrm{a}}=100~\mathrm{nm}$. Together, (A–B) show that disc-type devices achieve higher resistance values that scale continuously with $u_\text{a}$, while exhibiting reduced sensitivity to geometric variations compared to dome-type geometry.} \label{figure:fig2}
\end{figure}

\noindent Contrasting to these, in dome-type devices, current follows an approximately hemispherical path and the resistance can be expressed\cite{menzel2026device} as
$R_{\rm mush} \sim \frac{\rho_{\rm amor}}{2 \pi} \left(\frac{1}{r_{\rm BE}} - \frac{1}{u_\text{a}}\right)$.  A first observation is that a disc-type device includes a $1/t_{\rm PCM}$ prefactor, which introduces an additional design knob for resistance tuning through the PCM thickness, enabling large resistance values by using ultra-thin films.  A second important feature is that, in the limit of ultra-thin films
(typically $\leq 10~\mathrm{nm}$), the above behavior combines with the
inherent thickness-dependent properties of the phase-change material. In
particular, the crystalline and amorphous resistivities,
$\rho_\text{crys}$ and $\rho_\text{amor}$, are no longer well described by
constant bulk values, but instead increase nonlinearly as the film thickness
is reduced. This leads to further amplification in the resistance values across all phase configurations.  The third feature is the logarithmic dependence of resistance on the amorphous radius $u_\text{a}$. This allows for a gradual variation of resistance for RESET states, making it favorable for analog programming (see Figure~\ref{figure:fig2}A). \\

\noindent The fourth critical characteristic feature arises from a fabrication standpoint. Here, the first observation is that because the full-channel resistance scales with the geometric ratio $r_{\rm PCM}/r_{\rm BE}$, as the bottom-electrode radius decreases, the channel span can be simultaneously shrunk without compromising the memory window. The second observation is that the geometric variations enter as additive contributions in log-space ($\ln(r_{\rm PCM}) - \ln(r_{\rm BE})$), resulting in a natural compression of variability. This means variability in bottom-electrode has a lesser effect on the device's read-out performance. Equivalently, this reduced sensitivity is captured by the magnitude of the derivative $\left|\frac{\partial R}{\partial r_{\rm BE}}\right| \propto \frac{1}{r_{\rm BE}}$, thereby mitigating the impact of fabrication-induced variations (see Figure \ref{figure:fig2}B). 

\section*{Understanding the RESET Characteristics}  

\noindent Since switching in the devices is governed by heating, when a current pulse $I_\text{prog}$ of sufficiently high amplitude is applied to a PCM device, a significant portion of the phase change material heats up owing to dissipation of electrical power. This equates to $
\dot{Q} \propto P_\text{prog} = \gamma R_{\text{dynamic}} \times I_\text{prog}^2$, where \( \gamma \) defines the fraction of the heat dissipated in the cell, here assumed to be unity, and $R_{\text{dynamic}}$ is the resistance of cell when a portion of the phase-change material volume exists in the high electric-field regime. Therefore, $R_{\text{dynamic}} = R_\text{H} + R_{\text{ON}} + R_{\text{crys}}$, where $R_\text{H}$ is the heater resistance, $R_{\text{ON}}$ is the resistance of the phase-change material in the high-electric-field regime, and $R_{\text{crys}}$ is the resistance of the surrounding crystalline phase-change material. The ON-state resistance of the molten phase-change material region is given by

\begin{equation*}\label{eq:prog_RON}
R_{\text{ON}}(u_{\text{a}},T) = \left[\frac{\rho_{\text{melt,0}}}{2 \pi \times t_{\text{PCM}}} \times (\ln(u_{\text{a}})-\ln(r_{\text{BE}}))\right] \exp\left(\frac{E_{\text{melt}}}{k_{\text{b}}T}\right)
\end{equation*}

where $\rho_{\text{melt,0}}$ is the temperature-independent pre-exponential factor of the melt-state electrical resistivity. The temperature rise in the cell can be equated to $T_{\text{hot}} = T_{\text{amb}} + \gamma I_{\text{prog}}^2 R_{\text{dynamic}} \times R_\text{th}$, where, $T_{\text{amb}}$ is the ambient temperature, taken as \unit[300]{K}, and  $R_\text{th}$ is the thermal resistance for heat flow out of the hot spot. Since heat is predominantly removed through the electrodes, $R_\text{th}$ is modeled as the parallel combination of the corresponding thermal resistances. For a disc-type cell, these paths correspond to heat flow into the bottom electrode, lateral heat spreading through the crystalline phase-change material toward the surrounding top-electrode region, and vertical heat flow through the dielectric toward the top electrode.

\begin{equation*}\label{eq:prog_RTH_Tot}
\frac{1}{R_\text{th}(u_{\text{a}})} = \frac{1}{R_{\text{BE-Interface}}^{\text{th}}} + \frac{1}{R_{\text{TE-Interface}}^\text{th}(u_{\text{a}})} + \frac{1}{R_{\text{SE-Interface}}^\text{th}(u_{\text{a}})}
\end{equation*}

Considering the edge contacted disc-type geometry,
$R_{\text{BE}}^{\text{th}} = h_{\text{BE}}/(k_{\text{BE}}\pi r_{\text{BE}}^2)$,
$R_{\text{TE}}^{\text{th}} = h_{\text{oxide}}/(k_{\text{oxide}}\pi u_{\text{a}}^2)$, and
$R_{\text{SE}}^{\text{th}} =
\log(r_{\text{PCM}}/u_{\text{a}})/(2\pi k_{\text{crys}}t_{\text{PCM}})$.
Here, \(k_{\text{crys}}\), \(k_{\text{BE}}\), and \(k_{\text{oxide}}\) are the thermal conductivities of the crystalline PCM, bottom electrode, and oxide, respectively, while \(h_{\text{BE}}\) and \(h_{\text{oxide}}\) denote the heater height and dielectric thickness, respectively. Using the equations described above, the relationship between the hot spot size ($u_{\text{a}}$) and the corresponding programming current $I_\text{prog}$ can be expressed as\cite{BoniardiIEDM2014} $
I_{\text{prog}} = 
\sqrt{\frac{T_{\text{melt}} - T_{\text{amb}}}{\gamma \times R_{\text{dynamic}}(u_{\text{a}}) \times R_\text{th}(u_{\text{a}})}}$. With this expression, it thus becomes evident that the combination of high heat generation and reduced heat dissipation is crucial to efficient reset operation. \\

\begin{figure}[h!]
    \includegraphics[width=\textwidth]{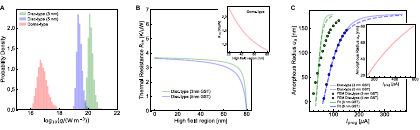}
    \caption{\textbf{RESET Characteristics}. (A) Comparison of power density between disc-type and dome-type geometries, illustrating heat generation efficiency in the devices. Disc-type cells exhibit improved power density. (B) Thermal resistance of  disc-type devices for various sizes of melt volumes. The inset shows a similar plot for a dome-type geometry. Notably, the absolute values of $R_\text{th}$ are increased, and its spatial dependency is reduced in disc-type devices. (C) Programming-current dependence of the phase-change region in disc-type devices with 3 and 5 nm GST. Solid curves show the analytical electrothermal model, symbols represent FEM results, and dashed curves are stretched-exponential fits. Enhanced self-heating in the disc geometry reduces the programming current required to extend the molten region. The inset shows the corresponding behavior for a dome-type geometry.} \label{fig:figure3}
\end{figure}

The first observation we make is the enhanced power density in disc-type devices compared to dome-type geometry. The volumetric Joule heating scales as $q = \mathbf{J} \cdot \mathbf{E}$, where $\mathbf{J}$ is the current density and $\mathbf{E}$ is the electric field. Crucially, the current density is enhanced in disc-type cells due to spatial confinement (see Figure~\ref{fig:figure3}A), leading to a substantially higher power density and as before an improved tolerance to variations in the bottom electrode heater. In addition, thickness reduction further amplifies the power density. Therefore, the efficiency of Joule heat generation within the phase-change film is high in disc-type device. \\

Furthermore, similar to the electrical behavior, the thermal properties are also expected to favorably improve in disc-type devices.\cite{aryana2021interface} In Figure~\ref{fig:figure3}B, we compute the effective thermal resistance of disc-type devices and compare it with that of a dome-type device. Notably, the thermal resistance is higher in the disc-type cells. In Figure~\ref{fig:figure3}C, we plot the analytically derived programming curves and compare them with FEM simulations and with the dome-type geometry, illustrating a notable reduction in programming current. This reduction results from both enhanced heat generation, which is predominantly a volume-dependent effect, and reduced heat dissipation, which is predominantly a surface-dependent effect. The resulting dependence of the phase-change radius on programming current can be conveniently parameterized by the stretched-exponential relation $
u_{\mathrm{a}}(I_{\mathrm{prog}})
=
r_{\mathrm{PCM}}
\exp\left[
-\ln\left(\frac{r_{\mathrm{PCM}}}{r_{\mathrm{BE}}}\right)
\left(\frac{I_{\mathrm{on}}}{I_{\mathrm{prog}}}\right)^n
\right]
\label{eq:ua_stretched_exp}$, as shown by the dashed fits in Figure~\ref{fig:figure3}C, where $I_{\mathrm{on}}$ denotes the onset current for phase change and $n$ determines the current dependence of the phase-change-region expansion. 

\section*{Understanding the SET Characteristics}  

\noindent We now turn to the threshold-switching characteristics, which are primarily governed by the electric field across the amorphous region and by the electro-thermal mechanisms described earlier. Threshold switching occurs when the local electric field becomes sufficiently large to induce a strong increase in the conductivity of the amorphous phase and, consequently, enhanced Joule heating~\cite{buckleyEvidenceCriticalFieldSwitching1974,Y2020legalloJPD}. \\

The electrical transport in the amorphous phase is strongly dependent on both temperature and electric field. This behavior is described using a three-dimensional two-center Poole--Frenkel transport model, whose parameters are extracted from temperature-dependent subthreshold $I$--$V$ measurements, as shown in \textcolor{black}{Supplementary Figure S3}. The corresponding local Joule-heating density is generally given by $ p_{\rm J}(\mathbf{r}) =
\mathbf{J}(\mathbf{r})\cdot\mathbf{E}(\mathbf{r}) =
\sigma\!\left(E,T\right)E^2$,  such that the total generated Joule power within the active amorphous region is $
P_{\rm J}
=
\int_{V_{\rm a}}
\sigma\!\left(E,T\right)E^2\,dV$. \\

Within an effective electro-thermal description, threshold switching is reached when the generated Joule power is sufficient to raise the active region to a characteristic switching temperature $T_{\rm S}$, taken here as \unit[400]{K}, such that

\begin{equation*}
P_{\rm J}\left(E_{\rm th},T_{\rm S}\right)
=
\frac{T_{\rm S}-T_{\rm amb}}{R_{\rm th}}
\end{equation*}

Because $\sigma(E,T)$ increases strongly with electric field, $E_{\rm th}$ must in general be determined self-consistently from this electro-thermal balance rather than from a field-independent conductivity approximation. Nevertheless, over the relatively narrow range of device geometries considered here, the switching field remains confined within a limited range, such that the variation of $V_{\rm th}$ is predominantly governed by the extent of the amorphous region $u_{\rm a}$. Furthermore, experiments reveal a pronounced thickness dependence of the activation energy $E_{\rm a}$, with an increase of approximately $\Delta E_{\rm a}\approx$ \unit[0.1]{eV}  in disc-type devices. An increase in $E_{\rm a}$ reduces the conductivity at a given field and temperature and therefore tends to increase the field required to reach the switching condition. In contrast, the enhanced thermal resistance of the disc-type geometry reduces the Joule power required to reach $T_{\rm S}$ and consequently lowers the required switching field. Crucially, these two effects compete, and the increase in $R_{\rm th}$ can compensate for, or even overcome, the increase in $E_{\rm a}$, resulting in a reduced $E_{\rm th}$ for the disc-type geometry. \\

In order to extend this field and geometry dependent voltage scaling, equations describing the dynamic Joule heating must be computed. For a RESET state, $R_{\text{device}} = R_0 \times \exp \left(\frac{E_\text{a} - \beta_{\text{eff}} \sqrt{V_{\text{device}}}}{K_B T}\right)$, where, $\beta_{\text{eff}} = \frac{\beta}{\sqrt{u_{\text{a}}}} \, , \,  E_\text{a} = E_\text{a,0}-\frac{a \times T^2}{b+T}$, $V_{\text{device}} = V_{\text{app}} \times \left( \frac{R_{\text{device}}}{R_{\text{device}}+R_{\text{heater}}+R_{\text{ser}}} \right)$. From the above, we can compute the current in the device, $I_{\text{device}} = \frac{V_{\text{device}}}{R_{\text{device}}}$. This allows us to dynamically capture the current in the cell, and couple it with thermal feedback comprising the effective thermal resistance we calculated in the previous section. 


\begin{equation*}\label{eq:TS_dT/dt}
\frac{\mathrm{d}T_\mathrm{amor}}{\mathrm{d}t} =  \frac{R_\mathrm{th}I_\mathrm{device}V_\mathrm{app}-\left( T_\mathrm{amor}-T_\mathrm{amb} \right)}{ R_\mathrm{th}C_\mathrm{th}} 
\end{equation*}

Here, $C^\text{th}$ is the thermal capacitance of the device. The temperature rise at time step in the cell can then be expressed as, $T_{\mathrm{amor}(t)} = T_{\mathrm{amor} (t-\Delta t)} +  \frac{\mathrm{d}T_{\mathrm{amor}}}{\mathrm{d}t}\mathrm{d}t$.\\




Using the same approach as for deriving the field profiles for read-out characteristics, the threshold voltages ($V_\mathrm{th}$) can be expressed as a function of the $u_\mathrm{a}$, 

\begin{equation*}
\mathbf{E}_{\rm amor}(r, V_0) = 
- \frac{V_0}{r} 
\left(
\frac{\sigma_{\rm crys}}{\sigma_{\rm crys} \log\frac{u_\text{a}}{r_{\rm TH}} + \sigma_{\rm amor} \log\frac{r_{\rm PCM}}{u_\text{a}}}
\right) \hat{\mathbf{e}}_r, 
\quad r_{\rm TH} < r < u_\text{a},
\end{equation*}

where $r$ is the radial coordinate, $\hat{\mathbf{e}}_r$ is the radial unit vector, $r_{\rm TH}$ is the radius of the top heater, $r_{\rm PCM}$ is the PCM disc radius, $\sigma_{\rm amor}$ and $\sigma_{\rm crys}$ are the resistivities of the amorphous and crystalline phases, respectively, and $V_0(t)$ is the applied voltage, assumed to vary slowly in time. The electric field in the amorphous region must reach a critical value $E_{\rm th}$, implying that a specific voltage $V_{\rm th}$ must be applied to the device, \begin{equation*}
V_{\rm th} = E_{\rm th} u_\text{a} \left(
\frac{\sigma_{\rm crys} \log\frac{u_\text{a}}{r_{\rm TH}} + \sigma_{\rm amor} \log\frac{r_{\rm PCM}}{u_\text{a}}}{\sigma_{\rm crys}}
\right).
\end{equation*}

For typical PCM materials, where $\sigma_{\rm amor} \ll \sigma_{\rm crys}$, this expression simplifies to $V_{\rm th} \approx E_{\rm th} \, u_\text{a} \, \log \frac{u_\text{a}}{r_{\rm TH}}$. $V_{\rm th}$ can be interpreted as the product of two contributions.  For $u_\text{a} \to r_{\rm BE}$, the expression reduces to $V_{\rm th} \sim E_{\rm th} (u_\text{a} - r_{\rm BE})$, leading to a linear dependence of $V_{\rm th}$ on $u_\text{a}$.   For large $u_\text{a}$, the dependence becomes log-linear, reflecting the logarithmic term in the expression. \\

\begin{figure}[h!]
    \includegraphics[width=0.65\textwidth]{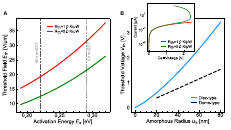}
    \caption{\textbf{Threshold Switching Characteristics}. (A) Comparison of threshold field for different thermal resistance values. The logarithmic field dependence indicates higher voltages are required for switching. The dotted vertical lines are experimental values of activation energies of GST, with and without spaital confinement. (B) Computed threshold voltage characteristics as a function of amorphous volume in the disc and dome-type device geometries, showing a difference in their dependencies. The inset show the full current-voltage behavior of a full RESET state, highlighting the improved thermal resistance to be a key property in disc-type devices.} \label{figure:fig4}
\end{figure}

In Figure~\ref{figure:fig4}A, we compare the critical threshold field as a function of $E_{\rm a}$ for varying values of the effective thermal resistance, $R_{\mathrm{th}}$, independent of device geometry. We note that $E_{\rm a}$ increases in thinner films of the phase-change material, which would, in isolation, translate to higher threshold voltages. However, in disc-type devices, the reduction in effective thickness is accompanied by a concomitant increase in $R_{\mathrm{th}}$. This provides a compensatory mechanism that is crucial when comparing threshold fields between dome-type devices (\unit[80]{nm}) and disc-type devices (\unit[5]{nm}). From a device operation perspective, this has important implications for the voltage required to trigger threshold switching. While the threshold voltages are inherently higher in disc-type devices (see Figure~\ref{figure:fig4}B), as experimentally validated\cite{syedDisc2025}, they are favorably reduced due to the increased $R_{\mathrm{th}}$ (see inset).

\section*{Understanding the Failure Characteristics}  

\noindent The failure of PCM devices is primarily governed by the degradation of the active volume during repeated electro-thermal cycling\cite{novielli2013atomic}. Failure typically arises from the permanent geometrical modification of the bottom electrode and stoichiometric shifts in material composition within the active switching volume. These factors in most device geometries are linked through phase-change segregation, where the radius of the bottom electrode $r_{BE}$ progressively increases as metallic elements migrate and deposit at the interface. This atomic redistribution can result in an altered stoichiometry with different electrical properties. \\

We note that the radial phase configuration of a disc-type device provides an intrinsic robustness against variations in its lateral dimensions, while remaining sensitive to variations in the PCM thickness and material properties. The total high-field resistance $R_{\rm dynamic}$, which is directly accessible experimentally, provides a useful measure of such device variations.\cite{Y2009Bipin} For a disc-type geometry, it is given by the sum of the molten and crystalline contributions,

\begin{equation*}
R_{\rm dynamic}
=
\frac{\rho_{\rm melt}}{2\pi t_{\rm PCM}}
\ln\left(\frac{u_a}{r_{\rm BE}}\right)
+
\frac{\rho_{\rm crys}}{2\pi t_{\rm PCM}}
\ln\left(\frac{r_{\rm PCM}}{u_a}\right)
\label{eq:rdynamic_radial}
\end{equation*}

As discussed in the RESET characteristics section and shown in Figure~\ref{fig:figure3}C, the programming-current dependence of the phase-change radius $u_a$ can be parameterized by a stretched-exponential relation. Substituting this relation into equation~\ref{eq:rdynamic_radial} gives $
R_{\rm dynamic}
=
\frac{\ln(r_{\rm PCM}/r_{\rm BE})}
{2\pi t_{\rm PCM}}
\left[
\rho_{\rm melt}
+
\left(\rho_{\rm crys}-\rho_{\rm melt}\right)
\left(\frac{I_{\rm on}}{I_{\rm prog}}\right)^n
\right]
\label{eq:rdynamic_endurance}
$. Equivalently, this relation can be written as

\begin{equation*}
R_{\rm dynamic}
=
R_{\rm melt}
+
\Delta R
\left(\frac{I_{\rm on}}{I_{\rm prog}}\right)^n
\label{eq:rdynamic_compact}
\end{equation*}

where

\begin{equation*}
R_{\rm melt}
=
\frac{\rho_{\rm melt}}
{2\pi t_{\rm PCM}}
\ln\left(\frac{r_{\rm PCM}}{r_{\rm BE}}\right),
\qquad
\Delta R
=
\frac{\rho_{\rm crys}-\rho_{\rm melt}}
{2\pi t_{\rm PCM}}
\ln\left(\frac{r_{\rm PCM}}{r_{\rm BE}}\right).
\end{equation*}

\begin{figure}[h!]
    \centering
    \includegraphics[width=0.85\textwidth]{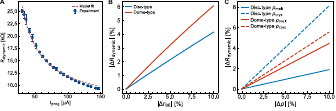}
    \caption{\textbf{Endurance characteristics.}
    (A) Experimental $R_{\rm dynamic}$ as a function of programming current for a disc-type device, together with the fit obtained from the analytical model.
    (B) Magnitude of the relative variation in $R_{\rm dynamic}$ resulting from variations in the bottom-electrode radius $r_{\rm BE}$ for disc- and dome-type phase configurations.
    (C) Magnitude of the relative variation in $R_{\rm dynamic}$ resulting from variations in the molten- and crystalline-state resistivities, $\rho_{\rm melt}$ and $\rho_{\rm crys}$, for the two device geometries. The geometrical and material variations serve as proxies for structural and compositional changes that may develop during extensive cyclic stressing.}
    \label{endurance}
\end{figure}

Figure~\ref{endurance}A compares this relation with the experimentally measured $R_{\rm dynamic}$ as a function of programming current for a disc-type device. Using experimentally extracted material resistivities and the phase-change onset current, the measured dependence can be described using $n$ as the only fitting parameter, providing an experimental connection between the programming-induced phase configuration and the measured electrical response. This model also highlights the origin of the geometric robustness of the disc-type phase configuration. Variations in the bottom-electrode radius enter explicitly only through the logarithmic factor $\ln(r_{\rm PCM}/r_{\rm BE})$, whereas the resistance scales directly as $t_{\rm PCM}^{-1}$. Consequently, variations in lateral dimensions are partially suppressed by the radial geometry, while variations in film thickness remain directly reflected in the device resistance. Figure~\ref{endurance}B shows the resulting resistance variation as a function of changes in $r_{\rm BE}$ and compares the disc-type geometry with a conventional dome-type phase configuration. \\

The same expression further separates the contributions of the molten- and crystalline-state resistivities. Defining $x
=
\left(\frac{I_{\rm on}}{I_{\rm prog}}\right)^n$,
the dynamic resistance can be expressed as
$R_{\rm dynamic}
=
\frac{\ln(r_{\rm PCM}/r_{\rm BE})}
{2\pi t_{\rm PCM}}
\left[
\rho_{\rm melt}(1-x)
+
\rho_{\rm crys}x
\right]$. The relative importance of the two material properties therefore depends on the programmed phase configuration. Close to the onset of phase change, $I_{\rm prog}\simeq I_{\rm on}$ such that $R_{\rm dynamic}$ is predominantly sensitive to $\rho_{\rm crys}$. At larger programming currents, $x$ decreases and the contribution of $\rho_{\rm melt}$ becomes progressively more important. Figure~\ref{endurance}C compares the sensitivity of disc- and dome-type devices to variations in both $\rho_{\rm melt}$ and $\rho_{\rm crys}$, representing possible compositional and structural changes induced during extensive cycling.\\

Another mechanism influencing device operability is the retention of the phase configuration. The key factors determining retention characteristics are the activation energy, $E_\text{barrier}$, a material property encompassing both nucleation and growth barriers, and a geometrical factor that accounts for the surface-to-volume ratio. In the experimentally observed case where $E_\text{barrier,dome} = E_\text{barrier,disc}$, and in a growth-dominated crystallization mode with growth velocity $v_\text{g}$, the confinement due to ultra-small thicknesses benefits disc-type device crystallization. For materials that crystallize heterogeneously from amorphous--crystalline interfaces, the time to full crystallization due to shrinkage of the amorphous volume follows $V_{\text{am,2D}}(t) \approx V_{\text{am,3D}}(t) \cdot \frac{3h}{2 \, (u_{a,0} - v_\text{g} t)}$. Specifically, the reduction in dimensionality ensures that the amorphous volume in disc-type device persists as the volumetric decay is moderated from cubic to quadratic dependence on the shrinking radius. However, in a crystallization mode dominated by bulk nucleation, or when additional interfaces promote nucleation, disc-type devices become more susceptible to SET failure.

\section*{Understanding the Scaling Properties}  

\begin{figure}[h!]
    \includegraphics[width=0.65\textwidth]{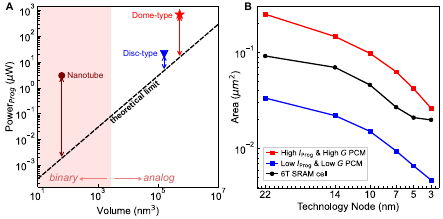}
    \caption{\textbf{Current Status and Future Scaling}. (A). Comparison of device geometries with the theoretical adiabatic programming limit. Notably, disc-type devices approach this limit most closely, while achieving the optimal volume for analog programming. (B) A plot illustrating the areal scaling of the compute unit cell with advancing CMOS technology nodes. The comparison includes two PCM device configurations and benchmarks them against an SRAM-based compute unit cell.} \label{Outlook}
\end{figure}

\noindent We now discuss key aspects of the disc-cell geometry and its scaling trends relevant for analog \gls{imc}. A primary goal is to approach the theoretical adiabatic limits of programming\cite{xiong2016towards}, while keeping the analog behavior. These, in turn, are set by fundamental material and geometrical properties. To quantify this limit, we estimate the minimum programming power required to induce the phase transition under ideal adiabatic conditions. This limit is determined by the energy required to heat the material to complete phase transition\cite{raoux2014phase}, $P_{\text{adiabatic}} = \frac{\rho \left( C_\text{p} \Delta T + L_{\text{fusion}} \right) V_\text{active}}{t_{\text{pulse}}}$. Here, $C_\text{p}$, $\Delta T$ and $L_{\text{fusion}}$ are material constants, and $V$ is the active volume. For analog behavior, the active volume must contain a sufficient number of grains to enable gradual and reproducible conductance modulation. Figure~\ref{Outlook}A shows that disc type cells can achieve a near optimal active volume for analog programming while progressively approaching the adiabatic limit. Further improvements are expected from continued scaling of the active volume as well as the contact areas. \\

Scaling down the film thickness provides the benefit of increased electrical resistivity. However, further reduction becomes primarily limited by the uniformity of deposition.  While industry-standard sputter deposition methods allow for thickness control on the order of $\approx 5~\mathrm{nm}$ across $300~\mathrm{mm}$ wafers, the relative film roughness and chemical inhomogeneity tend to increase for thinner films. In terms of thickness, the increased vulnerability arises from the inverse dependence on the device thickness, $t_\mathrm{PCM}$, such that $\frac{dR}{dt} \propto \frac{1}{t_\mathrm{PCM}^2}$. Therefore, for disc-type devices, conformal deposition techniques such as atomic layer deposition and chemical vapor deposition may therefore become more desirable. Scaling down the the contact area can be provisioned through reducing the heater radius. For the geometric dimensions of our devices, this reduction in contact area approaches \( \lim_{r \to h} \frac{r}{2h} = \frac{1}{2} \). Within this context, we distinguish between two contact geometries considered 
earlier: the planarized heater forming surface contact with the phase-change film, and a geometry in which the heater fully penetrates into the phase-change material, resulting in an edge contact with the phase-change material. We note that the absolute value of the contact area determines which geometry provides more favorable scaling. This, in turn, depends on the thickness of the phase-change film. For a fixed film thickness, the fully penetrating geometry is advantageous when \( r_{\text{BE}} < 2 t_{\text{PCM}} \), and vice versa.  As these aspects are optimized, the lowest achievable conductance state scales with the ratio \( r_{\text{PCM}} / r_{\text{BE}} \). Consequently, reducing the heater radius enables a proportional reduction in device footprint, naturally aligning with advances in lithographic scaling of minimum feature sizes. For example, as \( r_{\text{BE}} \) approaches current industry limits\cite{song202112nanoscale,Y2020arnaudIEDM} (currently \unit[3.5]{nm}), the projected trends can be compared with present-day devices.\\

In Figure~\ref{Outlook}B, we highlight the prospects of these scaling behavior. We project the current mushroom-type device demonstrated for \gls{imc} onto CMOS technology node scaling\cite{syed2023memory,Y2023legalloNatElec}. As expected, the integration density improves with smaller transistors capable of delivering higher current. However, the comparison shows that while this provides an avenue for scaling, it can only keep pace with the scaling trend of SRAM-based cells (shown in the black trace) at the smallest technology nodes\cite{singh2025design}. Note that SRAM-based macros are currently the dominant commercial implementation of \gls{imc}\cite{axelera_metis}. Disc type devices can break this constraint owing to their high programming efficiency, which reduces the number of required selectors, and their high resistance values, which allow integration closer to the transistors and thereby enable further downscaling. This is essential for achieving the integration density promised by analog \gls{imc} based on non volatile memory devices. Lastly, There is potential for further improvements from phase-change material itself\cite{sarwat2017materials}. Superlattice films\cite{Y2021kwonNanoletters,Y2021khanScience} and select doped-compositions \cite{redaelli2022material,ryu2008phase} can provide inherent confinement properties to improve the programming efficiency, as well as improve other characteristics, including endurance and retention. \\

In summary, we have detailed the underlying mechanisms that govern the behavior of disc-type phase change memory devices during read-out, RESET, SET switching, and endurance operations. These mechanisms are compared with analytically modeled dome-type phase configurations. Additionally, we have discussed the scaling characteristics of these devices, alluding to the pathway for high density PCM for computational memory applications. 

\section*{Acknowledgments}

\noindent This work was supported by the European Research Council Grant INFUSED (Grant No. 101222715) and by the IBM Research AI Hardware Center. We thank Andrea Cassini, Jesse Luchtenveld, Stephan Menzel and Timothy Philicelli for valuable technical discussions and contributions, and Vijay Narayanan for his management support. 

\section*{References}
\def\url#1{}
\bibliographystyle{naturemag}
\bibliography{References}

\end{document}